\RequirePackage{fix-cm}
\documentclass[smallextended]{svjour3}       
\smartqed  
\usepackage{amssymb}
\usepackage{amsmath}
\journalname{}
\begin{document}

\title{Improvement of a quantum broadcasting multiple blind signature scheme based on quantum teleportation 
}


\author{Wei Zhang \and Dao-Wen Qiu \and Xiang-Fu Zou }


\institute{W. Zhang  \and D.-W. Qiu   \at
              Department of Computer Science, Sun Yat-sen University, Guangzhou 510006, China \\
              The Guangdong Key Laboratory of Information Security Technology, Sun Yat-sen
University, Guangzhou 510006, China\\
              \email{issqdw@mail.sysu.edu.cn}\\
              \and
W. Zhang \at
              Department of Mathematics, Qiannan Normal College for Nationalities, Duyun 558000, China \\
              \and
              X.-F. Zou  \at
              School of Mathematics and Computational Science, Wuyi University, Jiangmen 529020, China \\
                            }

\maketitle

\begin{abstract}
 Recently, a broadcasting multiple blind signature scheme based on quantum teleportation has been proposed for the first time. It is claimed to have unconditional security and properties of quantum multiple signature and quantum blind signature. In this paper, we analyze the security of the protocol and show that each signatory can learn the signed message by a single particle measurement and the signed message can be modified at random by any attacker  according to the scheme. Furthermore, there are some participant attacks and external attacks existing in the scheme. Finally we present an improved scheme and show that it can resist all of the mentioned attacks. Additionally, the secret keys can be used again and again, making it more efficient and practical.

\end{abstract}

\section{Introduction}

\label{introduction}
Quantum signature is the counterpart in the quantum world of classical digital signature. Most classical digital signature schemes are based on public key cryptography which can be broken by Shor's algorithm \cite{1}. Quantum signature, which is based on the laws of quantum physics, can provide us unconditional security. Many different quantum signature models are proposed for different application demands, such as arbitrated quantum signature \cite{2,3,4,5,6,7}, quantum proxy signature \cite{8,9,10,11}, quantum group signature \cite{12,13,14,15}, quantum blind signature \cite{16,17,18} and quantum multiple signature \cite{19,20}.

A secure quantum signature scheme should satisfy the following two basic requirements: (1) No forgery. Specifically, the signature cannot be forged by any illegal signatory. (2) No disavowal. The signatory cannot disavow his signature and the receiver cannot disavow his receiving it. Furthermore, the receiver cannot disavow the integrity of the signature  \cite{4}.

As quantum cryptography has developed, many cryptanalysis of existing protocols have been presented \cite{21,22,23,24,25,26,27}. Some effective attack strategies also have been proposed to eavesdrop in the existing quantum cryptography protocols \cite{28}, such as intercept-resend attacks \cite{29}, entanglement swapping attacks \cite{30,31,32}, teleportation attacks \cite{33,34}, dense-coding attacks \cite{35,36,37}, channel-loss attacks \cite{38,39}, denial-of-service attacks \cite{40,41}, correlation$-$extractability attacks \cite{42,43,44}, Trojan horse attacks \cite{45,46,47,48}, participant attacks \cite{49} and collaborate attacks \cite{50}. Understanding these attacks is very important for designing quantum signature schemes with higher security. It also advances the research in quantum signature. Zou and Qiu  \cite{4} analyzed the arbitrated quantum signatures based on GHZ states and Bell states, finding that the receiver Bob can successfully reject the signature by disavowing its integrity. Then they proposed a new scheme by using a public board to fix this security loophole in which the entanglement was not needed any more.

Gao et al.  \cite{21} gave a perfect cryptanalysis on existing arbitrated quantum signature. They found that the signature can be forged by the receiver at will in almost all the existing AQS schemes and the sender can disavow the signature just by an intercept-resend method. Due to the existence of serious loopholes, it is imperative to reexamine the security of other quantum signature protocols.

Recently, a broadcasting multiple blind quantum signature scheme based on quantum teleportation has been proposed in Ref. \cite{51}. It is said to have the properties of both quantum multiple signature and quantum blind signature. Here we show that it is not a real blind signature because the signatory can get the content of the signed message. In addition, the signed message can be modified at random by any attacker. Moreover, there are some participant attacks and external attacks existing in the scheme. For instance, the message sender Alice can impersonate $U_{i}$ successfully as she can get the content of the signature and $U_{i}$'s secret key $K_{CU_{i}}$. Moreover, Alice can sign arbitrary message at will. The signature collector Charlie can counterfeit the signature optionally. With respect to the external attacks, the eavesdropper Eve can forge $U_{i}$'s signature at will without knowing the secret key $K_{CU_{i}}$.

All the attack strategies are described in detail and finally we present an improved scheme which can resist all the mentioned attacks. Meanwhile, since all the secret keys can be reused, it may greatly increase the scheme's efficiency and make it more practical.

The rest of this paper is organized as follows. First, in Section 2 we review the original protocol briefly. In Section 3 we present the security analysis of the original protocol and describe the attack strategies in detail. In Section 4 we present an improved scheme and analyze its security by showing that the improved one can resist all the attacks mentioned above and that the keys can be used again and again. In Section 5 a short conclusion is given and an issue worthy on further research is proposed.

\section{Review of the original protocol \cite{51}}

\label{Review of Yuan Tian's protocol}

The protocol involves the following four characters: (1) Alice is the message sender. (2)  $U_{i}$ is the $i$-th member of broadcasting multiple signatory. (3) Charlie is the signature collector. (4) Bob is the receiver and the verifier of the broadcasting multiple blind signature.

The scheme is composed of three parts: the initial phase, the individual blind signature generation and verification phase, and the combined multiple blind signature verification phase.

In this scheme, Alice sends $t$ copies of $n$-bit classical message $m$ to $t$ signatories $U_{i}$ $(i=1,2,\cdots, t)$ respectively, then $U_{i}$ signs the message $m$ to get the blind signature $S_{i}$ and sends $S_{i}$ to Charlie. Charlie collects and verifies these blind signatures, then he constructs a multiple signature and sends it to Bob. Finally, Bob verifies the multiple signature by confirming the message.

(1) Initial phase

(1.1) Alice transforms the classical message $m$ into $n$-bit as
\begin{eqnarray}
&& m= m(1)||m(2)|| \cdots ||m(j)||\cdots ||m(n),\\\nonumber
&& m(j)=0 \quad \text{or} \quad m(j)=1,\\\nonumber
&& j=1, 2, \cdots, n.
\end{eqnarray}

(1.2) Quantum key distribution

Alice shares a secret key $K_{AB}$ with Bob, a secret key $K_{AC}$ with Charlie, and secret keys $K_{AU_{i}}$ $(i=1,2,\cdots,t)$ with each signatory $U_{i}$, respectively. Bob shares a secret key $K_{BC}$ with Charlie, Charlie shares secret keys $K_{CU_{i}}$ $(i=1,2,\cdots,t)$ with each signatory $U_{i}$ respectively. To obtain unconditional security, all these keys are distributed via QKD protocols.

(1.3) Alice sends $E^{C}_{K_{AB}}(m)$ to Bob

Here $E^{C}$ means classical one-time pad algorithm,
\begin{eqnarray}
E^{C}_{K_{AB}}(m)=K_{AB}\oplus m.
\end{eqnarray}
 $E_{K}^{Q}$ in the later means quantum one-time pad
 \begin{eqnarray}
E^{Q}_{K}(|P\rangle)=\bigotimes_{i=1}^{n}\sigma_{x}^{k_{2i-1}}\sigma_{z}^{k_{2i}}|P_{i}\rangle,
\end{eqnarray}
$K$ is a secret key with $|K|= 2n$, $K_{i}$ is the $K$'s $i$-bit. $|P\rangle$ is an $n$-bit quantum message, $|P_{i}\rangle$ is its $i$-bit. $\sigma_{x}$ and $\sigma_{z}$ are two Pauli operators.

(2) The individual blind signature generation and verification phase

In this phase, we pick one of the signatory $U_{i}$ as the representative who signs the message.

(2.1) Message transformation

Assume that Alice is to send the message $m$. She prepares $n$-qubit state $|\psi(m)\rangle_{M}$ as
\begin{eqnarray}
&& |\psi(m)\rangle_{M}=\bigotimes^{n}_{j=1}|\psi(j)\rangle_{M},
\end{eqnarray}

where
\begin{equation}    |\psi(j)\rangle_{M} =
 \begin{cases}
    \frac{1}{\sqrt{2}}(|0\rangle_{M}+|1\rangle_{M})  &  \text{if  \quad $m(j)=1$} \\
    \frac{1}{\sqrt{2}}(|0\rangle_{M}-|1\rangle_{M})
        &  \text{if  \quad $m(j)=0$.}
 \end{cases}
 \end{equation}

(2.2) Quantum channel setup

Alice prepares $n$ EPR pairs. Each pair is denoted as

\begin{eqnarray}
|a(j)\rangle_{AC}=\frac{1}{\sqrt{2}}(|00\rangle_{AC}+|11\rangle_{AC}), j=1,2,3,\cdots,n+l.
\end{eqnarray}

(2.3) Signature Phase

(2.3.1) Alice picks up her $n$ EPR particles denoted as $\{|\varphi(N)\rangle_{A}\}$, i.e.
\begin{eqnarray}
&& \{|\varphi(N)\rangle_{A}\}= \{|\varphi(1)\rangle_{A}, |\varphi(2)\rangle_{A}, \cdots, |\varphi(j)\rangle_{A}, \cdots, |\varphi(n)\rangle_{A}\},
\end{eqnarray}
and the other $n$ EPR particles denoted as $\{|\varphi(N)\rangle_{C}\}$, i.e.
\begin{eqnarray}
&& \{|\varphi(N)\rangle_{C}\}= \{|\varphi(1)\rangle_{C}, |\varphi(2)\rangle_{C}, \cdots, |\varphi(j)\rangle_{C}, \cdots, |\varphi(n)\rangle_{C}\}.
\end{eqnarray}

(2.3.2) To distinguish each signatory, Alice creates a unique serial number which is denoted as $SN$ attaching to $\{|\varphi(N)\rangle_{A}\}$. Since $SN$ is a classical string, Alice transfers it to a quantum state sequence $|SN\rangle$ with the basis $B_{Z}=\{|0\rangle,|1\rangle\}$. Then she sends $E^{Q}_{K_{AU_{i}}}(|\psi(N)\rangle_{MA}, |SN\rangle)$ to $U_{i}$. Here
\begin{eqnarray}
&& |\psi\rangle_{MA}= \bigotimes ^{n}_{j=1}|\psi(j)\rangle_{M}\otimes|\varphi(j)\rangle_{A}.
\end{eqnarray}
After that, Alice sends $E^{Q}_{K_{AC}}(\{|\varphi(N)\rangle_{C}\}, |SN\rangle)$ to Charlie.

(2.3.3) $U_{i}$ decrypts $E^{Q}_{K_{AU_{i}}}(|\psi\rangle_{MA}, |SN\rangle)$ to get $|\psi\rangle_{MA}$ and $|SN\rangle$, then he performs Bell-basis measurement to get the outcomes $\{\beta_{MA}(j)|j=1, 2, \cdots, n\}$. Each $\beta_{MA}(j)=\beta_{kl}(k,l\in\{0, 1\})$ is expressed by 2-bit string $kl$ according to $|\beta_{kl}\rangle \mapsto kl$. Then he gets $S_{i}$ as

\begin{eqnarray}
&& S_{i}=\beta_{MA}(1)||\beta_{MA}(2)|| \cdots || \beta_{MA}(j) || \cdots || \beta_{MA}(n).
\end{eqnarray}

(2.3.4) $U_{i}$ sends $E^{C}_{K_{CU_{i}}}(S_{i}, SN)$ to Charlie.

(2.4) Verification Phase

(2.4.1) Charlie decrypts $E^{C}_{K_{CU_{i}}}(S_{i}, SN)$to get the signature $S_{i}$ and $SN$.

(2.4.2) According to $S_{i}$, $SN$ and quantum teleportation, Charlie performs one of the corresponding reverse transformation $(I,X,Y,Z)$ on each particle  $|\varphi(j)\rangle_{C}$ in his hand to get $|\psi'(j)\rangle_{C}$. He obtains $|\psi'(m)\rangle_{C}$ as
\begin{eqnarray}
|\psi'(m)\rangle_{C}= \bigotimes ^{n}_{j=1} |\psi'(j)\rangle_{C}.
\end{eqnarray}

(2.4.3) Charlie gets $m'$ by measuring each $|\psi'(j)\rangle$ in the basis of $\{\frac{1}{\sqrt{2}}(|0\rangle_{M}+|1\rangle_{M}), \frac{1}{\sqrt{2}}(|0\rangle_{M}-|1\rangle_{M})\}$. Then he sends $E^{C}_{K_{BC}}(m')$ to Bob.

(2.4.4) Bob decrypts $E^{C}_{K_{BC}}(m')$, $E^{C}_{K_{AB}}(m)$ by the secret keys $K_{BC}$ and $K_{AB}$ respectively and compares $m$ with $m'$. If they are the same, $S_{i}$ is accepted. Otherwise, it is rejected.

(3) The combined multiple signature generation and verification phase

(3.1) Charlie collects all individual signatures to generate the multiple signature $S=\{S_{i}|i=1, 2, \cdots, t\}$ and generates the message $\{m'_{i}|i=1, 2, \cdots, t\}$. If $m'_{i}$ is equal to $m'_{i+1}$ $(i=1, 2$, $\cdots$, $t-1)$, he confirms the message and sends $E^{C}_{K_{BC}}(m'_{1})$ to Bob. If it is not equal, the process is terminated.

(3.2) After Bob decrypts $E^{C}_{K_{BC}}(m'_{1})$ and $E^{C}_{K_{AB}}(m)$, he accepts $S$ if $m'_{1}$ is equal to $m$, otherwise he terminates the process.

\section{Cryptanalysis of the original protocol }

 In this section, we point out that there are some security loopholes in the scheme in Ref. [51] and describe the corresponding attack strategies in detail.

\subsection {Each $U_{i}$ can learn the signed message $m$}

The scheme is claimed to have properties of quantum blind signature so that the signatory cannot learn the signed message. Here we show that each signatory $U_{i}$ can get the message just by a single particle measurement.

Suppose Alice wants to send an $n$-bit classical message $m$ to get $U_{i}$'s signature, according to the scheme, she will transform it into $n$-qubit state $|\psi(m)\rangle_{M}$ according to Eq. (4) and Eq. (5). Because $\frac{1}{\sqrt{2}}(|0\rangle_{M}+|1\rangle_{M})$ and $\frac{1}{\sqrt{2}}(|0\rangle_{M}-|1\rangle_{M})$ are orthogonal to each other, they can form an orthonormal basis of the two dimensional Hilbert Space. When $U_{i}$ gets $K^{Q}_{AU_{i}}(|\psi(m)\rangle_{MA})$ from Alice in the signature phase, he can decrypt it and perform a single particle measurement in the basis of $\{\frac{1}{\sqrt{2}}(|0\rangle_{M}+|1\rangle_{M}), \frac{1}{\sqrt{2}}(|0\rangle_{M}-|1\rangle_{M})\}$ on the first state to get the message $m$, which has no effect on the following process. From this problem, we can see the original scheme is not a real blind signature scheme.

\subsection {Any attacker can modify Message $m$ at random}

Here we show the signed message $m$ can be modified at random through the intercept-resend method by any attacker, including participant attackers or external attackers.

In the original scheme, message $m$ and $m'$ are encrypted according to the one-time pad encryption algorithm during their transmission. Any attacker can intercept $E^{C}_{K_{AB}}(m)$ and resend $E^{C}_{K_{AB}}(m)\oplus m_{0}$ to Bob in Step (1.3). According to the scheme, Bob will get $m \oplus m_{0}$ instead of $m$, here $m_{0}$ is an arbitrary 2$n$-bit random binary string. At the same time, he intercepts $E^{C}_{K_{BC}}(m')$ and resends $E^{C}_{K_{BC}}(m')\oplus m_{0}$ in Step (2.4.3). According to the scheme, $m \oplus m_{0}$ can pass the following verification process. Because $m_{0}$ is arbitrary, $m$ can be modified at random by any attacker through intercept-resend method.

\subsection { Alice's attack}

To illustrate Alice's attack, here take a 1-bit message $m(j)$ to make a demonstration.

\subsubsection {Alice can get the signature}

Suppose Alice sends the message $m(j)$ to get $U_{i}$'s blind signature $S_{i}(j)$. From the scheme, we can see that $U_{i}$ signs $m(j)$ by measuring $|\psi(j)\rangle_{MA}$ in the Bell basis, which is sent from Alice in Step(2.3.3). Alice can get $S_{i}(j)=\beta_{kl}$ by measuring $|\psi(j)\rangle_{MA}$ on Bell basis and recording the outcome $\beta_{kl}$ before she sends it to $U_{i}$. Instead, Alice sends the two particle state $|\beta_{kl}\rangle_{MA}$ to $U_{i}$. Then $U_{i}$'s measurement outcome is $\beta_{kl}$. Then Alice can get each $S_{i}(j)$.

\subsubsection {Alice can get $U_{i}$'s secret key $K_{CU_{i}}$}

It has been illustrated that Alice can get each $S_{i}(j)$, then Alice can get $U_{i}$'s signature $S_{i}$. Alice can intercept $E^{C}_{K_{CU_{i}}}(S_{i})||SN$ when it is sent from $U_{i}$ to Charlie in Step (2.3.4). Because Alice knows $S_{i}$, she can extract $U_{i}$'s secret key $K_{CU_{i}}$ by adding $S_{i}$ to $E^{C}_{K_{CU_{i}}}(S_{i})$ as $K_{CU_{i}}= S_{i} \oplus E^{C}_{K_{CU_{i}}}(S_{i})$. Then she resends $E^{C}_{K_{CU_{i}}}(S_{i})||SN$ to Charlie. All of these cannot be discovered.

\subsubsection {Alice can sign the message at will}

We can see that Alice can completely replace the signatory $U_{i}$ to sign the message. In order to illustrate Alice can sign arbitrary message at will, we will  demonstrate the quantum teleportation process of the above protocol as follows:

Suppose that the particle $M$ carry a $1$-bit classical information $m(j)$ and the state of particle $M$ are denoted as
\begin{align}
&& |\psi(j)\rangle_{M}=\frac{1}{\sqrt{2}}(|0\rangle_{M}+d|1\rangle_{M}),\quad d=\pm 1.
\end{align}

The EPR pairs shared between Alice and Charlie are denoted as
\begin{align}
&& |a(j)\rangle_{AC}=\frac{1}{\sqrt{2}}(|00\rangle_{AC}+|11\rangle_{AC}).
\end{align}

The two states are combined to form a three particle state $|\Phi(j)\rangle$ as

\begin{eqnarray}
&&|\Phi(j)\rangle=|\psi(j)\rangle_{M}\otimes|a(j)\rangle_{AC}\\\nonumber
&&\quad \quad \quad =(\frac{|0\rangle_{M}+d|1\rangle_{M}}{\sqrt{2}})(\frac{|00\rangle_{AC}+|11\rangle_{AC}}{\sqrt{2}})\\\nonumber
&&\quad \quad \quad =\frac{1}{2}[|\beta_{00}\rangle_{MA}(\frac{|0\rangle_{C}+d|1\rangle_{C}}{\sqrt{2}})
+|\beta_{01}\rangle_{MA}(\frac{|1\rangle_{C}+d|0\rangle_{C}}{\sqrt{2}})\\\nonumber
&&\quad \quad \quad \quad +|\beta_{10}\rangle_{MA}(\frac{|0\rangle_{C}-d|1\rangle_{C}}{\sqrt{2}})
+|\beta_{11}\rangle_{MA}(\frac{|1\rangle_{C}-d|0\rangle_{C}}{\sqrt{2}})],
\end {eqnarray}

where
\begin{eqnarray}
&& |\beta_{00}\rangle_{MA}=\frac{|00\rangle_{MA}+|11\rangle_{MA}}{\sqrt{2}},\\
&& |\beta_{01}\rangle_{MA}=\frac{|01\rangle_{MA}+|10\rangle_{MA}}{\sqrt{2}},\\
&& |\beta_{10}\rangle_{MA}=\frac{|00\rangle_{MA}-|11\rangle_{MA}}{\sqrt{2}},\\
\text{and} &&|\beta_{11}\rangle_{MA}=\frac{|01\rangle_{MA}-|10\rangle_{MA}}{\sqrt{2}}.
\end {eqnarray}

From Eq. (14), we can see if the measurement outcome is $\beta_{00}$, the state of the particle $C$ is just the information state $|\psi(j)\rangle_{M}$.  Then we take operation $I$ on the state of $C$. If the measurement outcome is $\beta_{01}$, then we perform operation $X$ on $C$ to recover it to the information state. If the outcomes are $\beta_{10}$ and $\beta_{11}$, then we take the operation $Z$ and $Y$ respectively.

Here, we show Alice can modify the signature $S_{i}(j)$ at random as follows:  When Alice prepares $|\psi(j)\rangle_{M}$ and $|a(j)\rangle_{AC}$ in Step (2.1) and Step (2.2), she does not send $|\varphi(j)\rangle_{C}$ to Charlie and $|\psi(j)\rangle_{MA}$ to $U_{i}$ immediately. Instead, she performs a Bell-basis measurement on $|\psi(j)\rangle_{MA}$ to get the outcome $\beta_{kl}$ and she sends another Bell state $|\beta_{k'l'}\rangle_{MA}$ to $U_{i}$. Then the signature $S_{i}(j)=\beta_{kl}$ has been changed into $S'_{i}(j)=\beta_{k'l'}$. In order to make sure $S_{i}'(j)$ can pass the verification, Alice performs a corresponding operator $V$ on $|\varphi(j)\rangle_{C}$ before sending it to Charlie. Alice can derive the corresponding operator $V$ according to Eq. (14). Assume $|\psi(j)\rangle_{M}$ is teleported from Alice to Charlie and the measurement outcome is $\beta_{kl}$. Also using this equation, Charlie will performs a Pauli operator $V_{1}$ on $|\varphi(j)\rangle_{C}$ to make sure
\begin{eqnarray}
V_{1}|\varphi(j)\rangle_{C}\equiv |\psi(j)\rangle_{M}.
\end {eqnarray}
In other words, the particle $C$ is in the state
\begin{eqnarray}
|\varphi(j)\rangle_{C}\equiv V_{1}|\psi(j)\rangle_{M}.
\end {eqnarray}
Here $A\equiv B$ means A is equivalent to B except for a global phase. Alice performs the corresponding $V$ on $|\varphi(j)\rangle_{C}$, so
\begin{eqnarray}
 |\varphi(j)\rangle_{C} \equiv V^{\dag}V_{1}|\psi(j)\rangle_{M}.
\end {eqnarray}
Here $V^{\dag}$ is the conjugate transpose of $V$. When $S_{i}(j)$ is changed into $S_{i}'(j)$,  Charlie will take another Pauli operator $V_{2}$ on $|\varphi(j)\rangle_{C}$ to return it to the information state $|\psi(j)\rangle_{M}$, then
\begin{eqnarray}
|\varphi(j)\rangle_{C}\equiv V_{2}V^{\dag}V_{1}|\psi(j)\rangle_{M}\equiv |\psi(j)\rangle_{M}.
\end {eqnarray}
From Eq. (22), we can conclude that
\begin{eqnarray}
V_{2}V^{\dag}V_{1}= I \quad \text{or} \quad V= V_{1}V_{2}.
\end {eqnarray}

We take a simple example to make an illustration. Suppose Alice get the measurement outcome $\beta_{00}$, but she sends $|\beta_{01}\rangle_{AM}$ to $U_{i}$, according to Eq. (14), $V_{1}=I$, $V_{2}=X$, then $V=X$ according to Eq. (23). We list Alice's attack strategies in Table 1.

\begin{table}[h]
\center
\begin{tabular}{|r|ccc|}
\hline
\  $S_{i}(j)=\beta_{kl}\rightarrow S'_{i}(j)=\beta_{k'l'}$ & $V_{1}$ & $V_{2}$ & $V$  \\\hline
$S_{i}(j)=\beta_{00}\rightarrow S'_{i}(j)=\beta_{01}$ & $I$ & $X$ & $X$  \\
$S_{i}(j)=\beta_{00}\rightarrow S'_{i}(j)=\beta_{10}$ & $I$ & $Z$ & $Z$  \\
$S_{i}(j)=\beta_{00}\rightarrow S'_{i}(j)=\beta_{11}$ & $I$ & $Y$ & $Y$  \\
$S_{i}(j)=\beta_{01}\rightarrow S'_{i}(j)=\beta_{00}$ & $X$ & $I$ & $X$  \\
$S_{i}(j)=\beta_{01}\rightarrow S'_{i}(j)=\beta_{10}$ & $X$ & $Z$ & $Y$  \\
$S_{i}(j)=\beta_{01}\rightarrow S'_{i}(j)=\beta_{11}$ & $X$ & $Y$ & $Z$  \\
$S_{i}(j)=\beta_{10}\rightarrow S'_{i}(j)=\beta_{00}$ & $Z$ & $I$ & $Z$  \\
$S_{i}(j)=\beta_{10}\rightarrow S'_{i}(j)=\beta_{01}$ & $Z$ & $X$ & $Y$  \\
$S_{i}(j)=\beta_{10}\rightarrow S'_{i}(j)=\beta_{11}$ & $Z$ & $Y$ & $X$  \\
$S_{i}(j)=\beta_{11}\rightarrow S'_{i}(j)=\beta_{00}$ & $Y$ & $I$ & $Y$  \\
$S_{i}(j)=\beta_{11}\rightarrow S'_{i}(j)=\beta_{01}$ & $Y$ & $X$ & $Z$  \\
$S_{i}(j)=\beta_{11}\rightarrow S'_{i}(j)=\beta_{10}$ & $Y$ & $Z$ & $X$  \\\hline
\end{tabular}
\caption{Alice's attack strategies: First, Alice measures $|\psi(j)\rangle_{MA}$ to get $S_{i}(j)=\beta_{kl}$, but she sends $|\beta_{k'l'}\rangle_{MA}$ to $U_{i}$ instead. Then the signature $S_{i}(j)$ has been changed into $S'_{i}(j)$. At the same time, Alice performs a corresponding unitary operation $V$ on $|\varphi(j)\rangle_{C}$ before sending it to Charlie.}

\end{table}

\subsection { Charlie's attack}

In the original scheme, Charlie can also attack the program by modifying the signature $S$ at will.

Charlie is the signature collector whose duty is to collect all the individual signature $S_{i}(i=1, 2, \cdots, t)$ and extract $m'_{i}$ by first recovering each $|\varphi(j)\rangle_{C}$ to $|\psi'(j)\rangle_{C}$ according to Eq. (14) and then measuring $|\psi'(m)\rangle_{C}$ in the basis of $\{\frac{1}{\sqrt{2}}(|0\rangle_{M}+|1\rangle_{M}), \frac{1}{\sqrt{2}}(|0\rangle_{M}-|1\rangle_{M})\}$. Charlie can modify the signature $S$ into arbitrary $S'$ and keep the message $m'_{1}$ unchanged after confirming the message. Because Bob just verifies whether $m$ is equal to $m'_{1}$ or not, $S'$ can pass the verification without being discovered.

\subsection {  Eavesdropper Eve's forgery attack}

In Ref. [51], it is declared that the eavesdropper Eve can't forge $U_{i}$'s signature on the assumption that she can get $U_{i}$'s secret key $K_{CU_{i}}$ because of the quantum teleportation. Here we show that Eve can forge $U_{i}$'s signature at will even though she knows nothing about the $U_{i}$'s key $K_{CU_{i}}$.

Here we take a 1-bit message $m(j)$ to make a demonstration. This is an imcomplete message $m(j)$ whose signature is $S_{i}(j)$. Eve replaced $S_{i}(j)$ with another $S'_{i}(j)$ when it is sent from $U_{i}$ to Charlie. Then Charlie will recover the message according to $S'_{i}(j)$ based on teleportation. Suppose the signature $S_{i}(j)$ is $\beta_{00}$. It is changed into $S'_{i}(j)=\beta_{01}$ under Eve's attack. We take an illustration as follows:

(1) Without Eve's attack

Suppose the signature $S_{i}(j)$ is $\beta_{00}$ and Charlie's particle $C$ is in the state
\begin{align}
|\varphi(j)\rangle_{C}=\frac{1}{\sqrt{2}}(|0\rangle_{C}+d|1\rangle_{C}), d=\pm 1.
\end{align}
Then Charlie will perform $I$ on his particle to recover it to the information state $|\psi'(j)\rangle_{C}$ according to Eq. (14), here
\begin{align}
|\psi'(j)\rangle_{C}=I|\varphi(j)\rangle_{C}=\frac{1}{\sqrt{2}}(|0\rangle_{C}+d|1\rangle_{C}).
\end{align}
 After that Charlie measures it and extracts the message $m'(j)$ as
 \begin{equation}
 m'(j) =
 \begin{cases}
    1  &  \text{if  \quad $d=1$} \\
    0  &  \text{if  \quad $d=0$}.
 \end{cases}
 \end{equation}

(2) With Eve's attack

The signature is tampered with $S'_{i}(j)=\beta_{01}$ with Eve's attack, Charlie will perform $X$ on his particle $C$ which is still in the state of Eq. (24). Then the state of $C$ will be

\begin{align}
|\psi'(j)\rangle_{C}=X|\varphi(j)\rangle_{C}=\frac{1}{\sqrt{2}}(|1\rangle_{C}+d|0\rangle_{C}).
\end{align}
After that, Charlie measures $|\psi'(j)\rangle_{C}$ to extract the message $m''(j)$ as
 \begin{equation}
 m''(j) =
 \begin{cases}
    1  &  \text{if  \quad $d=1$} \\
    0  &  \text{if  \quad $d=0$}.
 \end{cases}
 \end{equation}

From Eq. (26) and Eq. (28), we can see Charlie will get the same messages, i.e., $m''(j)=m'(j)$. See the list of all the cases in Table 2.
\begin{table}[h]
\center
\begin{tabular}{|r|c|}
\hline
\  $S_{i}(j)=\beta_{kl}\rightarrow S'_{i}(j)=\beta_{k'l'}$ & $m''(j)$ and $m'(j)$  \\\hline
$S_{i}(j)=\beta_{00}\rightarrow S'_{i}(j)=\beta_{01}$ & $m''(j)=m'(j)$ \\
$S_{i}(j)=\beta_{00}\rightarrow S'_{i}(j)=\beta_{10}$ & $m''(j)=m'(j)\oplus 1$  \\
$S_{i}(j)=\beta_{00}\rightarrow S'_{i}(j)=\beta_{11}$ & $m''(j)=m'(j)\oplus 1$ \\
$S_{i}(j)=\beta_{01}\rightarrow S'_{i}(j)=\beta_{00}$ & $m''(j)=m'(j)$  \\
$S_{i}(j)=\beta_{01}\rightarrow S'_{i}(j)=\beta_{10}$ & $m''(j)=m'(j)\oplus 1$  \\
$S_{i}(j)=\beta_{01}\rightarrow S'_{i}(j)=\beta_{11}$ & $m''(j)=m'(j)\oplus 1$   \\
$S_{i}(j)=\beta_{10}\rightarrow S'_{i}(j)=\beta_{00}$ & $m''(j)=m'(j)\oplus 1$  \\
$S_{i}(j)=\beta_{10}\rightarrow S'_{i}(j)=\beta_{01}$ & $m''(j)=m'(j)\oplus 1$   \\
$S_{i}(j)=\beta_{10}\rightarrow S'_{i}(j)=\beta_{11}$ & $m''(j)=m'(j)$  \\
$S_{i}(j)=\beta_{11}\rightarrow S'_{i}(j)=\beta_{00}$ & $m''(j)=m'(j)\oplus 1$   \\
$S_{i}(j)=\beta_{11}\rightarrow S'_{i}(j)=\beta_{01}$ & $m''(j)=m'(j)\oplus 1$  \\
$S_{i}(j)=\beta_{11}\rightarrow S'_{i}(j)=\beta_{10}$ & $m''(j)=m'(j)$ \\\hline
\end{tabular}
\caption{The relation between $m'(j)$ and $m''(j)$ under the circumstance that the signature $S_{i}(j)$ is tampered with by $S'_{i}(j)$ under Eve's attack.}
\end{table}

From the second column of Table 2, we can see $\beta_{00}$ and $\beta_{01}$ are interchangeable and so is the $\beta_{10}$ and $\beta_{11}$. Specifically, when Eve tampered with $S_{i}(j)=\beta_{00}(\beta_{10})$ by $S'_{i}(j)=\beta_{01}(\beta_{11})$ or vice versa, Charlie will extract the same message. Precisely, $m'(j)$ is equal to $m''(j)$. Accordingly, $S'_{i}(j)$ can always pass the verification. Eve's other modification of the signature is not interchangeable. We can see all the other cases get different message $m''(j)\neq m'(j)$, but they all satisfy $m''(j)=m'(j)\oplus 1$ where $\oplus$ is modulo 2 addition.

From Table 2, we can make a law of the message and its corresponding signature as follows:

If the signature $S_{i}(j)=\beta_{kl}$ is changed into  $S'_{i}(j)=\beta_{k'l'}$, then their corresponding messages $m'(j)$ and $m''(j)$ ($k,k',l,l'\in\{0,1\} \quad j=1, 2, \cdots, n$) will satisfy
 \begin{equation}
 m''(j) =
 \begin{cases}
    m'(j)  &  \text{if  \quad $k=k'$} \\
    m'(j)\oplus 1  &  \text{if  \quad $k\neq k'$}.
 \end{cases}
 \end{equation}

Next we show Eve can forge each signature $S_{i}$ by the intercept-resend method. Eve can intercept $E^{C}_{K_{CU_{i}}}(S_{i})$ when it is sent from $U_{i}$ to Charlie. She adds a 2$n$-bit binary string
\begin{align}
l=i_{1}i_{2}\cdots i_{2n}
\end{align}
to $E^{C}_{K_{CU_{i}}}(S_{i})$ and sends it to Charlie. Then Charlie will get
\begin{align}
S'_{i}=S_{i}\oplus l.
\end{align}
Charlie will recover the information $m''$ based on $S'_{i}$ according to the teleportation rather than $S_{i}$. Then Charlie will get
\begin{align}
m''=m'\oplus l',
\end{align}
where
\begin{align}
l'=j_{1}j_{2}\cdots j_{n}.
\end{align}
According to Eq. (29), $l'$ must satisfy
 \begin{equation}
 j_{k} =
 \begin{cases}
    0  &  \text{if  \quad $i_{2k-1}=0$} \\
    1  &  \text{if  \quad $i_{2k-1}=1$}.
 \end{cases}
 \end{equation}
At the same time, Eve intercepts $E^{C}_{K_{AB}}(m)$ in the Step (1.3) and resends $E^{C}_{K_{AB}}(m)\oplus l'$ to Bob. Then Bob will get $m\oplus l'$ instead of $m$. $S'_{i}$ will be accepted for the signature of $m\oplus l'$. We can see that Eve's forgery attack can get successful.

\section {  An improved scheme}

Here we present an improved scheme and show it can resist all the attacks mentioned above. Also the secret keys can be reused which can provide efficiency and practicality.

Before we present the improved scheme, it is necessary to introduce the QOTP algorithm it uses. Suppose a quantum message
\begin{align}
|P\rangle=\bigotimes^{n}_{j=1}|P_{j}\rangle
\end{align}
is composed of $n$ qubits
\begin{align}
|P_{j}\rangle=\alpha_{j}|0\rangle+\beta_{j}|1\rangle,
\end{align}
where
\begin{align}
|\alpha_{j}|^{2}+|\beta_{j}|^{2}=1
\end{align}
and the encryption key $K\in \{0, 1\}^{4n}$. The QOTP encryption $E_{K}$ used in this scheme on the quantum message can be described as
\begin{eqnarray}
&& E_{K}(|P\rangle)=\bigotimes^{n}_{j=1} \sigma^{K_{4j}}_{x}\sigma^{K_{4j-1}}_{z}T\sigma^{K_{4j-2}}_{x}\sigma^{K_{4j-3}}_{z}|P_{j}\rangle
\end{eqnarray}
where
\begin{align}
T=\frac{i}{\sqrt{3}}(\sigma_{x}-\sigma_{y}+\sigma_{z}).
\end{align}

This QOTP encryption $E_{K}$ is the improved one introduced in Ref. [52] for the first time. The assistant operator $T$ can make sure the encrypted message can not be forged in the scheme. Distinctly, for arbitrary message $|P\rangle$, there are no non-identity unitary $V$ and unitary $U$ such that
\begin{align}
E^{\dag}_{K}VE_{K}|P\rangle \equiv U |P\rangle.
\end{align}

In order to make the secret keys reusable in the improved scheme, we use a one-way hash function here \cite{7}:
 \begin{eqnarray}
&&H(x): \{0, 1\}^{*}\longrightarrow \{0, 1\}^{4n}.
\end{eqnarray}

This scheme also contains four factors: (1) Alice is the message sender. (2)  $U_{i}$ ($i=1, 2, \cdots, t$) is the $i$-th member of broadcasting multiple signatory. (3) Charlie is the signature collector. (4) Bob is the receiver and the verifier of the broadcasting multiple blind signature.

The improved scheme is also composed of three parts: the initial phase, the individual blind signature generation and verification phase, and the combined multiple blind signature verification phase.

(1) Initial Phase

(1.1) Quantum key distribution

Alice shares the secret key $K_{AB}$ with Bob, $K_{AC}$ with Charlie, and $K_{AU_{i}}$ $(i=1, 2, \cdots, t)$ with each signatory $U_{i}$; Bob shares a secret key $K_{BC}$ with Charlie; Charlie shares secret keys $K_{CU_{i}}$ $(i=1,2,\cdots,t)$ with each signatory $U_{i}$.  All the secret keys are 4$n$-bit. To obtain unconditional security, all these keys are distributed via QKD protocols.

(1.2) Message concealing and message transformation

 Alice gets $m'=m\oplus r$ where $m$ is an $n$-bit classical message  and $r$  an $n$-bit random binary string. Alice transforms the classical message $m'$ into $n$-qubit state
\begin{eqnarray}
|\psi(m')\rangle_{M}=\bigotimes^{n}_{j=1}|\psi(j)\rangle_{M},
\end{eqnarray}
where
\begin{equation}
 |\psi(j)\rangle_{M} =
 \begin{cases}
    b|0\rangle_{M}+c|1\rangle_{M} \quad  &  \text{if  \quad $m'(j)=1$} \\
    c|0\rangle_{M}-b|1\rangle_{M} \quad  &  \text{if  \quad $m'(j)=0$},
 \end{cases}
 \end{equation}
 b, c are different real constants.

(1.3) Alice sends the message $m$ to Bob

 Alice transforms $m$ into $|m\rangle$ as
\begin{eqnarray}
|m\rangle=\bigotimes^{n}_{j=1}|m(j)\rangle,
\end{eqnarray}
where
\begin{equation}
 |m(j)\rangle  =
 \begin{cases}
    |0\rangle &  \text{if  \quad $m(j)=0$} \\
    |1\rangle &  \text{if  \quad $m(j)=1$}.
 \end{cases}
 \end{equation}
Alice randomly chooses a 4$n$-bit binary sequence $r_{0}$, computes $H(K_{AB}||r_{0})$, and sends $[E_{H(K_{AB}||r_{0})}(|m\rangle)]\otimes |r_{0}\rangle$ to Bob where
\begin{eqnarray}
&& |r_{0}\rangle=\bigotimes^{4n}_{j=1}|r_{0}(j)\rangle,
\end{eqnarray}
and
\begin{equation}
 |r_{0}(j)\rangle  =
 \begin{cases}
    |0\rangle &  \text{if  \quad $r_{0}(j)=0$} \\
    |1\rangle &  \text{if  \quad $r_{0}(j)=1$}.
 \end{cases}
 \end{equation}

Bob extracts $r_{0}$ by measuring each particle of $|r_{0}\rangle$ in the basis $\{|0\rangle, |1\rangle\}$. Then he can compute $H(K_{AB}||r_{0})$ and decrypt $E_{H(K_{AB}||r_{0})}(|m\rangle)$ to get $|m\rangle$. After that he can get $m$ by performing a measurement in the basis $\{|0\rangle, |1\rangle\}$.

(1.4) Alice sends $r$ to Charlie

 $|r\rangle$ is generated as Eq. (46) and Eq. (47). Further more, all the $|r_{k}\rangle$'s generation in the rest of the scheme is always the same. Alice sends $E_{H(K_{AC}||r_{1})}(|r\rangle)\otimes |r_{1}\rangle$ to Charlie. Then Charlie extracts $r$ by performing a measurement in the computational basis.

(2) The individual blind signature generation and verification phase

In this phase, we pick one of the signatory $U_{i}$ as the representative who signs the message.

(2.1) Quantum channel setup

 Charlie prepares $n+l$ pairs of EPR particles denoted as $\{|a(1)\rangle_{U_{i}C}, $ $|a(2)\rangle_{U_{i}C}, $ $ \cdots,  $ $|a(n+l)\rangle_{U_{i}C}\}$, where $|a(j)\rangle_{U_{i}C}=\frac{1}{\sqrt{2}}(|00\rangle_{U_{i}C}+|11\rangle_{U_{i}C}),$$ j=1, 2, \cdots, n+l$. Then he sends the first particle to $U_{i}$ and keeps the other himself for every EPR pair. After $U_{i}$ receives all particles, Charlie randomly chooses $l$ particles to perform a measurement randomly in the basis of $\{|0\rangle, |1\rangle\}$ or $\{\frac{1}{\sqrt{2}}(|0\rangle+|1\rangle), \frac{1}{\sqrt{2}}(|0\rangle-|1\rangle)\}$ and reports the position of the particles that he has measured and the basis that he has chosen to $U_{i}$. $U_{i}$ takes the same measurement on the corresponding particles and compares the measurement outcomes with Charlie. If there is no error, the channel is considered to be safe. Otherwise, they abandon the quantum channel and set it up again.

(2.2) Signature Phase

(2.2.1) Alice sends the information quantum state $|\psi(m')\rangle_{M}$ to $U_{i}$

Alice randomly chooses a 4$n$-bit sequence $r_{2}$ and computes $H(K_{AU_{i}}||r_{2})$. Then she sends $[E_{H(K_{AU_{i}}||r_{2})}(|\psi(m')\rangle_{M})]$ $\otimes |r_{2}\rangle$ to $U_{i}$.

(2.2.2) $U_{i}$ signs the message $m$

 $U_{i}$ decrypts $[E_{H(K_{AU_{i}}||r_{2})}(|\psi(m')\rangle_{M})]$ to get $|\psi(m')\rangle_{M}$ by first extracting $r_{2}$ and then computing $H(K_{AU_{i}}||r_{2})$. He generates $\{|\psi(j)\rangle_{MU_{i}}| j=1, 2, \cdots, n\}$ by combining each $|\psi(j)\rangle_{M}$ with his EPR particle. Then $U_{i}$ performs a Bell basis measurement on $\{|\psi(j)\rangle_{MU_{i}}| j=1, 2, \cdots, n\}$ to get the outcomes $\{\beta_{MU_{i}}(j)| j=1, 2, \cdots, n\}$. According to Eq. (14), $\beta_{MU_{i}}(j)$ is an Bell state $|\beta_{kl}\rangle$ which can be expressed in 2-bit classical string according to $|\beta_{kl}\rangle \longrightarrow kl, k,l \in \{0, 1\}$.  By introducing a 4$n$-bit random binary string $R_{i}$, $U_{i}$ gets the blind signature $S_{i}$ of $m'$ as
\begin{eqnarray}
S_{i}=(\beta_{i}\oplus K_{CU_{i}})|| H[(\beta_{i}\oplus K_{CU_{i}})||R_{i}]
\end{eqnarray}
where
\begin{eqnarray}
\beta_{i}=\beta_{MU_{i}}(1)||\beta_{MU_{i}}(2)||\cdots||\beta_{MU_{i}}(j)||\cdots||\beta_{MU_{i}}(n).
\end{eqnarray}

(2.2.3) $U_{i}$ sends $|S_{i}\rangle$ to Charlie

 $U_{i}$ transforms $S_{i}$ into quantum state $|S_{i}\rangle$ as

\begin{eqnarray}
|S_{i}\rangle=\bigotimes^{6n}_{j=1}|S_{i}(j)\rangle
\end{eqnarray}
where
\begin{equation}
 |S_{i}(j)\rangle  =
 \begin{cases}
    |0\rangle &  \text{if  \quad $S_{i}(j)=0$} \\
    |1\rangle &  \text{if  \quad $S_{i}(j)=1$}.
 \end{cases}
 \end{equation}

$U_{i}$ randomly chooses a $6$ dimensional 4$n$-bit string vector $r_{3}=(r^{1}_{3}, r^{2}_{3}, r^{3}_{3}, r^{4}_{3}, $ $r^{5}_{3}, r^{6}_{3})$ and computes $H(K_{CU_{i}}||r_{3})$. Then he sends $[E_{H(K_{CU_{i}}||r_{3})}$ $(|S_{i}\rangle)] \otimes |r_{3}\rangle$ to Charlie where
\begin{eqnarray}
H(K_{CU_{i}}||r_{3})=H(K_{CU_{i}}||r^{1}_{3})||H(K_{CU_{i}}||r^{2}_{3})||H(K_{CU_{i}}||r^{3}_{3})\\\nonumber
                     ||H(K_{CU_{i}}||r^{4}_{3})||H(K_{CU_{i}}||r^{5}_{3})||H(K_{CU_{i}}||r^{6}_{3})
\end{eqnarray}
and
\begin{eqnarray}
|r_{3}\rangle=|r^{1}_{3}\rangle \otimes|r^{2}_{3}\rangle \otimes|r^{3}_{3}\rangle \otimes|r^{4}_{3}\rangle \otimes|r^{5}_{3}\rangle \otimes|r^{6}_{3}\rangle.
\end{eqnarray}

(2.3) Verification Phase

(2.3.1) Charlie decrypts $E_{H(K_{CU_{i}}||r_{3})}(|S_{i}\rangle)$ to get $|S_{i}\rangle$, then he can get $S'_{i}$ by performing a measurement in computational basis. After that he further gets $\beta'_{i}$ based on $K_{CU_{i}}$ according to Eq. (48). Here
\begin{eqnarray}
S'_{i}=(\beta'_{i}\oplus K_{CU_{i}})|| [H[(\beta_{i}\oplus K_{CU_{i}})||R_{i}]]'
\end{eqnarray}
If there is no incorrection happened in the transmission and measurement, $\beta'_{i}$ and $S'_{i}$ will be equal to $\beta_{i}$ and $S_{i}$ respectively.

(2.3.2) According to $\beta'_{i}$ and quantum teleportation in Eq. (14), Charlie performs one of the corresponding reverse transformation $(I,X,Y,Z)$ on each particle in his hand. He obtains the states of these particle denoted as $\{ |\psi'(j)\rangle_{C}| j=1, 2, \cdots, n\}$, which carry information of message $m''$.

(2.3.3) Charlie gets $m''$ by measuring each $|\psi'(j)\rangle_{C}, j=1, 2, \cdots, n$ in the basis of $\{b|0\rangle_{M}+c|1\rangle_{M}, c|0\rangle_{M}-b|1\rangle_{M}\}$. Then he computes $m^{*}=m''\oplus r$ and sends $[E_{H(K_{BC}||r_{4})}(|m^{*}\rangle)]\otimes |r_{4}\rangle$ to Bob.

(2.3.4) Bob decrypts $E_{H(K_{BC}||r_{4})}(|m^{*}\rangle)$ to get $|m^{*}\rangle$ and further gets $m^{*}$ by performing a measurement in the basis of $\{|0\rangle, |1\rangle\}$. Then he compares $m^{*}$ with $m$. If they are not equal, $S_{i}$ is rejected. Otherwise, Bob informs Charlie and $U_{i}$ to announce $S'_{i}$  and $R_{i}$ on the public board, respectively. Then he computes $H[(\beta'_{i}\oplus K_{CU_{i}})||R_{i}]$ and compares it to $[H[(\beta_{i}\oplus K_{CU_{i}})||R_{i}]]'$. If they are the same, $S_{i}$ is accepted. Otherwise, $S_{i}$ is considered to be compromised and it is rejected.

(3) The combined multiple signature generation and verification phase

(3.1) Charlie generates the message sequence $\{m^{*}_{i}| i=1, 2, \cdots, t\}$ and collects all individual signature $\{S'_{i}| i=1, 2, \cdots, t\}$. If $m^{*}_{i}$ is equals to $m^{*}_{i+1}$, $(i=1,2,\cdots,t-1)$, he confirms  the message and generates the multiple signature $S=\{S'_{i}| i=1, 2, \cdots, t\}$. If not, the process is terminated. After Charlie confirms the message, he sends $[E_{H(K_{BC}||r_{5})}(|m^{*}_{1}\rangle)]\otimes |r_{5}\rangle$ to Bob.

(3.2) Bob decrypts $E_{H(K_{BC}||r_{5})}(|m^{*}_{1}\rangle)$ to get $|m^{*}_{1}\rangle$ and further gets $m^{*}_{1}$ by performing a measurement in the basis of $\{|0\rangle, |1\rangle\}$. Then he compares $m^{*}_{1}$ with $m$. If they are not equal, $S$ is rejected. Otherwise, Bob informs Charlie to announce $S$ and each $U_{i} (i=1, 2, \cdots, t)$ to announce $R_{i}$ on the public board, respectively. Then he computes $F$ and compares it to $F'$ where
\begin{eqnarray}
&& F= \{H[(\beta'_{i}\oplus K_{CU_{i}})||R_{i^{*}}], i, i^{*}=1, 2, \cdots,  t\}\\
&& F'= \{[H[(\beta_{i}\oplus K_{CU_{i}})||R_{i}]]', i=1, 2, \cdots,  t\}.
\end{eqnarray}
If $F'\subseteq F$, $S$ is accepted. If not, $S$ is rejected.

\par
Let's use Figure 1 to illustrate our quantum signature model as follows:
\\
\par

\begin{picture}(150,145)
\put(45,140){Signature Phase}
\put(205,140){Verification Phase }

\setlength{\unitlength}{1.5cm}
\put(0.5,1.3){\framebox(0.6,0.5){Alice}}

\put(2.5,0.1){\framebox(0.6,0.4){U$_{t}$}}
\put(2.5,0.8){\framebox(0.6,0.4){$\vdots$}}
\put(2.5,1.8){\framebox(0.6,0.4){U$_{2}$}}
\put(2.5,2.6){\framebox(0.6,0.4){U$_{1}$}}

\put(4.5,1.3){\framebox(0.8,0.5){Charlie}}

\put(6.5,1.3){\framebox(0.6,0.5){Bob}}

\put(0.1,-0.2){\dashbox{0.2}(3.5,3.4){}}
\put(4.2,-0.2){\dashbox{0.2}(3.1,3.4){}}
\put(0.1,2.9){$|\psi(m')\rangle_{M}$ in Step(2.2.1)}
\put(1.4,1.77){\vector(0,3){1.02}}
\put(1.2,1.88){\vector(0,3){0.92}}
\put(1.6,1.22){\vector(0,3){1.57}}
\put(1.8,0.78){\vector(0,3){2.01}}
\put(2,1.55){$|r\rangle$ in Step(1.4)}
\put(3.4,0.1){$|m\rangle$ in Step(1.3)}
\put(5.05,1.75){$|m_{1}^{\ast}\rangle$ in Step(2.3.3)}
\put(3.8,0.81){\line(0,3){1.99}}
\put(4.0,1.3){\line(0,3){1.5}}
\put(4.28,1.7){\line(0,3){1.09}}
\put(4.40,1.81){\line(0,3){0.99}}
\put(3.8,2.8){\vector(1,0){1.3}}
\put(5.1,2.7){$|S_{i}\rangle$ in Step(2.2.3)}
\put(3.9,0.5){\line(0,3){1.74}}
\put(4.1,0.5){\line(0,3){1.30}}
\put(4.28,0.5){\line(0,3){0.82}}
\put(4.40,0.5){\line(0,3){0.72}}
\put(3.9,0.5){\vector(1,0){1.2}}
\put(5.1,0.4){$|\varphi(j)\rangle_{U_{i}}$ in Step(2.1)}
\put(0.1,2.8){\line(1,0){2.3}}

\put(1.1,1.8){\vector(4,3){1.4}}
\put(1.1,1.3){\vector(4,-3){1.4}}
\put(1.1,1.7){\vector(4,1){1.4}}
\put(1.1,1.35){\vector(4,-1){1.4}}
\put(1.1,1.5){\vector(1,0){3.4}}

\put(3.1,2.8){\vector(4,-3){1.4}}
\put(3.1,2.0){\vector(4,-1){1.4}}
\put(3.1,1.0){\vector(3,1){1.4}}
\put(3.1,0.3){\vector(4,3){1.4}}
\put(4.5,1.8){\vector(-4,3){1.4}}
\put(4.5,1.7){\vector(-4,1){1.4}}
\put(4.5,1.3){\vector(-4,-3){1.4}}
\put(4.5,1.4){\vector(-3,-1){1.4}}

\put(5.3,1.5){\vector(1,0){1.2}}

\put(0.8,1.3){\vector(0,-1){1.3}}
\put(0.8,0.0){\vector(1,0){6.0}}
\put(6.8,0.0){\vector(0,1){1.3}}
\end{picture}

\vskip 2mm
\vskip 2mm
\vskip 2mm
\vskip 2mm
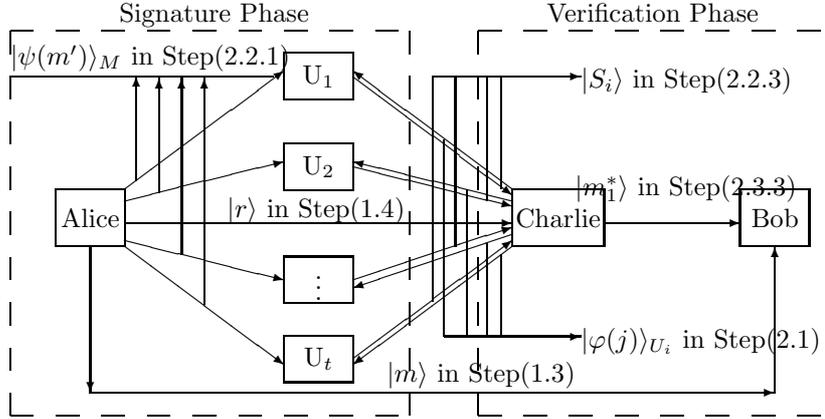
\centerline{\textbf{Figure 1.} The Quantum Signature Model of The Improved Scheme}
\vskip 2mm
\par

Next, we list the improvements of our new scheme compared to the original one:

(1) Introducing the improved QOTP encryption algorithm.

(2) Bringing in a hash function to authenticate the originality of the signature.

(3) The secret keys become reusable by introducing some random strings.

(4) Bringing in public boards.

(5) Classical message is concealed before turning into quantum message. Meanwhile, the transformation method in the improved scheme is according to Eq. (43) rather than Eq. (5) in the original one.

(6) The entangled quantum channel between Charlie and each $U_{i}(i=1, 2, \cdots, t)$ is set up by Charlie rather than Alice. At the same time, a channel checking process is added to make sure it is secure.

(7) Classical message from Alice to Bob is transmitted through quantum method in Step (1.3).

\section { Cryptanalysis of the improved scheme}

In this section, we present the security analysis of the improved scheme. we show there is no disavowal and forgery in the improved scheme. Meanwhile, we also point out the signatory cannot learn the signed message and the signed message cannot be modified by attackers in the improved scheme.

\subsection{No disavowal}

\subsubsection{ Each signatory $U_{i}$ cannot disavow his signature $S_{i}$ }

From Eq. (48), we can see that because each $S_{i}$ contains $U_{i}$'s secret key $K_{CU_{i}}$ in the improved scheme, $U_{i}$ cannot disavow his signature $S_{i}$. Meanwhile, $U_{i}$ cannot disavow $S_{i}$ by the intercept-resend method mentioned in Ref. \cite{21}. Because in Step (2.3.4) Charlie announces $S_{i}$ on the public board instead of sending it to Bob, which is only for reading on the public board, $U_{i}$ can't disavow his signature by intercept-resend method.

\subsubsection{ The receiver Bob cannot disavow the signature }

In the improved scheme, the signature $S$ is announced on the public board by Charlie when Bob informs him $m=m^{*}_{1}$ in Step (3.2) so that everyone can witness Bob has received the signature, so Bob can't disavow his receiving the signature. Also, Bob can't disavow the integrity of the signature by claiming $m\neq m^{*}_{1}$ as in Ref. \cite{4}. Assume $m= m^{*}_{1}$ and Bob lies to claim $m\neq m^{*}_{1}$ for his own benefit in Step (3.2). We can ask Alice , Charlie and Bob to public announce the message respectively. Then the dishonest behavior of Bob can be catught according to the voting rule, on the assumption that there is no collaborate attack.

\subsection{No forgery}

According to Eq. (48), each $S_{i}$ is composed of $\beta_{i}$, $K_{CU_{i}}$ and $R_{i}$. Here we show there is no participant forgery and external forgery in the improved scheme.

\subsubsection{ Alice cannot forge the signature}

In the improved scheme, Charlie prepares the EPR pairs $\{|a(j)\rangle_{CU_{i}}$ $| j=1, 2, \cdots, n+l\}$, Alice just prepares the information state $|\psi(m')\rangle_{M}$ so that she cannot get each $\beta_{i}$ by measuring each $|\psi(j)\rangle_{MU_{i}}$ directly. Next we show Alice cannot get $\beta_{i}$ by intercept-resend method either. Assume Alice intercepts each $|\varphi(j)\rangle_{U_{i}}$ and resends the measurement outcome $|\beta_{kl}\rangle_{MU_{i}}$ to $U_{i}$. Because Charlie and $U_{i}$ performs a checking process in Step (2.1), Alice's intercept-resend attack will be discovered. According to the improved scheme, Alice cannot get $\beta_{i}$. $K_{CU_{i}}$ is shared between Charlie and $U_{i}$ via QKD protocol so that Alice has no chance to get it. Moreover, $R_{i}$ is chosen by $U_{i}$ randomly and it is not acquired by anyone else until it is announced on the public board so that Alice cannot get it in the signing phase. Therefore, Alice cannot forge the signature.

\subsubsection{ Charlie cannot forge the signature}

Charlie, the signature collector who can get each $S'_{i}$ and the secret key $K_{CU_{i}}$ in the improved scheme, is considered to be most likely to forge the signature successfully. Here we show he cannot forge the signature either. Charlie can modify the signature at random and keep the message unchanged when he has confirmed the message, which has no influence on the following message comparison. Charlie cannot learn each $U_{i}$'s random string $R_{i}$. Since he does not know how to modify $H'[(\beta_{i}\oplus K_{CU_{i}})||R_{i}]$ to fit his modification deterministically, Charlie's forgery attack can definitely be discovered in Step (3.2).

\subsubsection{ Bob cannot forge the signature }

Bob is the receiver of the scheme, he cannot get the signature $S$ until Charlie announced it on the public board. Therefore, the only way Bob can modify the signature is to perform a unitary operator $V$ on $E_{H(K_{CU_{i}}||r_{3})}$ $(|S_{i}\rangle)$ in Step (2.2.3). According to Eq. (40), Bob's modification cannot follow his will, then it will be definitely discovered in the verification process. As a consequence, Bob cannot forge the signature.

\subsubsection{ The eavesdropper Eve cannot forge the signature}

In the improved scheme, the classical bit $m'(j)$ is transformed into quantum state according to Eq. (42) and Eq. (43). Here we take a 1-bit message $m'(j)$ to illustrate any of Eve's modification on $S_{i}(j)$ can be detected.

Supposed $S_{i}(j)=\beta_{01}$ is replaced with $S'_{i}(j)=\beta_{00}$ by Eve, the corresponding message are $m'(j)$ and $m''(j)$ respectively. We present this case as follows:

(1) Without Eve's attack:

(1.1) Assume Alice prepares the state $|\psi(j)\rangle_{M}=b|0\rangle_{M}+c|1\rangle_{M}$ and its signature is $ S_{i}(j)=\beta_{01}$, according to the teleportation, Charlie's particle will be in the state $|\varphi(j)\rangle_{C}= b|1\rangle_{C}+c|0\rangle_{C}$. Then Charlie performs operation $X$ on $|\varphi(j)\rangle_{C}$ to get the state $|\psi'(j)\rangle_{C}=b|0\rangle_{C}+c|1\rangle_{C}$. After that Charlie performs a measurement in the basis of $\{b|0\rangle_{M}+c|1\rangle_{M}, c|0\rangle_{M}-b|1\rangle_{M}\}$) to extract $m'(j)=1$.

(1.2) Assume Alice prepares the state $|\psi(j)\rangle_{M}=(c|0\rangle_{M}-b|1\rangle_{M})$ and its signature is $ S_{i}(j)=\beta_{01}$, according to the teleportation, Charlie's particle will be in the state $|\varphi(j)\rangle_{C}= c|1\rangle_{C}-b|0\rangle_{C}$. Then Charlie performs operation $X$ on $|\varphi(j)\rangle_{C}$ to get the state  $|\psi'(j)\rangle_{C}=c|0\rangle_{C}-b|1\rangle_{C}$. After that Charlie performs a measurement in the basis of $\{b|0\rangle_{M}+c|1\rangle_{M}, c|0\rangle_{M}-b|1\rangle_{M}\}$) to extract $m'(j)=0$.

(2) With Eve's attack:

(2.1) Assume Alice prepares the state $|\psi(j)\rangle_{M}=b|0\rangle_{M}+c|1\rangle_{M}$ and Eve replaces $S_{i}(j)=\beta_{01}$ with $S'_{i}(j)=\beta_{00}$,  then Charlie's particle is still in the state $|\varphi(j)\rangle_{C}= b|1\rangle_{C}+c|0\rangle_{C}$ as Eve's attack has no effect on it. According to the teleportation, Charlie will perform operation $I$ on $|\varphi(j)\rangle_{C}$ to get $|\psi'(j)\rangle_{C}=b|1\rangle_{C}+c|0\rangle_{C}$. After that Charlie performs a measurement in the basis of $\{b|0\rangle_{M}+c|1\rangle_{M}, c|0\rangle_{M}-b|1\rangle_{M}\}$) to extract $m''(j)$, it will have a probability of  $4b^{2}c^{2}$ to get $m''(j)=1$ and a probability of $c^{4}+b^{4}-2b^{2}c^{2}$ to get $m''(j)=0$.

(2.2) Assume Alice prepares the state $|\psi(j)\rangle_{M}=(c|0\rangle_{M}-b|1\rangle_{M})$ and Eve replaces $S_{i}(j)=\beta_{01}$ with $S'_{i}(j)=\beta_{00}$, then Charlie's particle is still in the state $|\varphi(j)\rangle_{C}= c|1\rangle_{C}-b|0\rangle_{C}$. According to the teleportation, Charlie will perform operation $I$ on $|\varphi(j)\rangle_{C}$ to get $|\psi'(j)\rangle_{C}= c|1\rangle_{C}-b|0\rangle_{C}$. After that Charlie performs a measurement in the basis of $\{b|0\rangle_{M}+c|1\rangle_{M}, c|0\rangle_{M}-b|1\rangle_{M}\}$ to extract $m''(j)$, it will have a probability of  $c^{4}+b^{4}-2b^{2}c^{2}$ to get $m''(j)=1$ and a probability of $4b^{2}c^{2}$ to get $m''(j)=0$.

From (1) and (2), we can see that Eve's modification of the signature can be discovered in Step (2.3.4) with a non-zero probability. Other cases can be presented similarly. We can see Eve cannot forge the signature.

\subsection{ Each $U_{i}$ cannot learn the signed message}

In the improved scheme, the classical message $m$ is turned into $m'=m\oplus r$ before it is transformed into quantum states. If $U_{i}$ performs a measurement in the basis of $\{b|0\rangle_{M}+c|1\rangle_{M}, c|0\rangle_{M}-b|1\rangle_{M}\}$ on the information quantum sequence $\{|\psi(j)\rangle_{M}| j=1, 2, \cdots, n\}$, he will just get $m'$ and has no chance to get the signed message $m$. If he wants to learn $m$, he has to know $r$ which will be sent from Alice to Charlie in Step (1.4). Because $r$ is turned into $|r\rangle$ and encrypted by $K_{AC}$ according to the quantum one-time pad algorithm during its transmission, $U_{i}$ cannot get the random string $r$ without the key $K_{AC}$. Therefore, we know $U_{i}$ can't learn the signed message $m$.

\subsection{ The signed message $m$ cannot be modified}
In the improved scheme, the signed message $m$ is turned into quantum state $|m\rangle$ according to Eq. (44) and Eq. (45) before it is sent from Alice to Bob in Step (1.3). It is encrypted by $H(K_{AB}||r_{0})$ according to Eq. (38). Here we show any attacker without the key $K_{AB}$ cannot modified the message $m$ by the intercept-resend method. Suppose the attacker wants to modify the signed message $m$. He intercepts the $[E_{H(K_{AB}||r_{0})}(|m\rangle)]\otimes |r_{0}\rangle$. Because he does not get the secret key $K_{AB}$, the attacker cannot decrypt it directly. Then he can choose to perform a unitary operator $V$ on $[E_{H(K_{AB}||r_{0})}(|m\rangle)]$ and send $V[E_{H(K_{AB}||r_{0})}(|m\rangle)]\otimes |r_{0}\rangle$ to Bob. In order to make sure this modification can pass the verification, there must exist a non-identity unitary operator $U$ to satisfy Eq. (40). Because $|m\rangle$ is from classical message $m$ according to Eq. (44) and Eq. (45), here $U$ can be restricted to Pauli operators. Exactly
\begin{eqnarray}
&& V[E_{H(K_{AB}||r_{0})}(|m\rangle)]\equiv E_{H(K_{AB}||r_{0})}(U|m\rangle)\\\nonumber
&& U=\bigotimes^{n}_{j=1}Q_{j}\\\nonumber
&& Q_{j}\in \{I, \sigma_{x}, \sigma_{y}, \sigma_{z}\}
\end{eqnarray}

The question the attacker has to face now is whether there exist such non-identity unitary operators $U$ and $V$ to satisfy Eq. (57). Unfortunately, it is pointed out that there doesn't exist such $U$ and $V$ for any message $|m\rangle$ in Ref. [52]. Then the modification mentioned above will be discovered in the verification process.

Next, we show Bob can modify the signed message $m$ at random in the improved scheme, but we can rebute it by the voting rule when this dispute takes place. Because Bob has got $K_{AB}$ and $K_{BC}$, according to the improved scheme, he can modify $m$ at random and make this modification can pass the verification. When the dispute on the message $m$ happens between Alice and Bob, we can ask Alice, Bob and Charlie to public their message and arbitrate it according to the voting rule on the assumption that they are all just loyal to themself. From above, we can see the signed message cannot be modified in the improved scheme.

\section {  Conclusion.}

In this paper, we first gave a security analysis on the quantum broadcasting multiple blind signature scheme based on teleportation in Ref. \cite{51}, which has recently been proposed. We point out that there are some security loopholes in the protocol and describe the attack strategies in detail. Then we present an improved scheme by introducing hash function, public board, and the improved QOTP encryption algorithm proposed in Ref. \cite{52}. After that, we show the improved scheme can resist all the mentioned attacks and that the secret keys can be reusable by bringing in some random strings. The improved scheme is more practical and secure. It will have foreseeable applications to E-payment system, E-business, and E-government.

The improved quantum broadcasting multiple blind signature can only sign classical message. So it is worthwhile for us to designing a scheme for quantum messages in the future.


\begin{thebibliography}{}
%
%

\bibitem{1}Shor, P.: Polynomial-time algorithms for prime factorization and discrete logarithms on a quantum computer. SIAM J. Comput. 26(5), 1484-1509 (1997)

\bibitem{2} Zeng G, Keitel C H.: Arbitrated quantum-signature scheme. Phys. Rev. A. 65(4), 042312 (2002)

\bibitem{3} Li, Q, Chan, W.H., Long, D.Y.: Arbitrated quantum signature scheme using Bell states. Phys. Rev. A. 79(5), 054307 (2009)

\bibitem{4} Zou X, Qiu D.: Security analysis and improvements of arbitrated quantum signature schemes. Phys. Rev. A. 82(4), 042325 (2010)

\bibitem{5} Li Q, Li C, Long D, et al.: Efficient arbitrated quantum signature and its proof of security. Quantum Inf. Process. 12(7), 2427-2439 (2013)

\bibitem{6} Luo Y P, Hwang T.: Arbitrated quantum signature of classical messages without using authenticated classical channels. Quantum Inf. Process. 13(1), 113-120 ( 2014)

\bibitem{7} Yu C.H, Guo G.D, Lin S.: Arbitrated quantum signature scheme based on reusable key. Sci. Ch. Phys.  Mech.  Astron.  57(11), 2079-2085 (2014)

\bibitem{8} Yin X R, Ma W P, Liu W Y.: Quantum proxy group signature scheme with ¦Ö-type entangled states. Int. J. Quantum Inform. 10, 1250041 (2012)

\bibitem{9} Wang T Y, Wei Z L.: One-time proxy signature based on quantum cryptography. Quantum Inf. Process. 11(2), 455-463 (2012)

\bibitem{10} Wen X, Chen Y, Fang J.: An inter-bank E-payment protocol based on quantum proxy blind signature. Quantum Inf. Process. 12(1), 549-558 (2013)

\bibitem{11} Cao H J, Huang J, Yu Y F, et al.: A Quantum Proxy Signature Scheme Based on Genuine Five-qubit Entangled State. Int. J. Theor. Phys. 53(9), 3095-3100 (2014)

\bibitem{12}  Wen X, Tian Y, Ji L, et al.: A group signature scheme based on quantum teleportation. Phys. Scr. 81(5), 055001 (2010)

\bibitem{13} Xiaojun W. :An E-payment system based on quantum group signature.  Phys. Scr. 82(6), 065403 (2010)

\bibitem{14} Xu R, Huang L, Yang W, et al.: Quantum group blind signature scheme without entanglement. Opt. Commun. 284(14), 3654-3658 (2011)

\bibitem{15}  Zhang K, Song T, Zuo H, et al.: A secure quantum group signature scheme based on Bell states.  Phys. Scr. 87(4), 045012 (2013)
\bibitem{16} Qi S, Zheng H, Qiaoyan W, et al.: Quantum blind signature based on two-state vector formalism. Opt. Commun. 283(21), 4408-4410 (2010)

\bibitem{17} Yin X R, Ma W P, Liu W Y. :A blind quantum signature scheme with ¦Ö-type entangled states. Int. J. Theor. Phys. 51(2), 455-461 (2012)

\bibitem{18}  Lin T S, Chen Y, Chang T H, et al.: Quantum blind signature based on quantum circuit. Nanotechnology (IEEE-NANO), 2014 IEEE 14th International Conference on. IEEE, 868-872 (2014)

\bibitem{19} Wen X J, Liu Y, Sun Y.: Quantum multi-signature protocol based on teleportation. Zeitschrift fur Naturforschung A. 62(3/4), 147 (2007)

\bibitem{20} Wen X, Liu Y.: A realizable quantum sequential multi-signature scheme. Dianzi Xuebao(Acta Electronica Sinica). 35(6), 1079-1083 (2007)

\bibitem{21} Gao F, Qin S J, Guo F Z, et al.: Cryptanalysis of the arbitrated quantum signature protocols. Phys. Rev. A. 84(2), 022344 (2011)

\bibitem{22}Choi J W, Chang K Y, Hong D.: Security problem on arbitrated quantum signature schemes. Phys. Rev. A. 84(6), 062330 (2011)

\bibitem{23} Zuo H, Zhang K, Song T.: Security analysis of quantum multi-signature protocol based on teleportation. Quantum Inf. Process. 12(7), 2343-2353 (2013)

\bibitem{24} Kejia Z, Dan L, Qi S.: Security of the arbitrated quantum signature protocols revisited. Phys. Scr. 89(1), 015102 (2014)

\bibitem{25} Yang C W, Luo Y P, Hwang T.: Forgery attack on one-time proxy signature and the improvement. Quantum Inf. Process. 13(9), 2007-2016 (2014)

\bibitem{26} Liu Z H, Chen H W, Wang D, et al.: Cryptanalysis and improvement of three-particle deterministic secure and high bit-rate direct quantum communication protocol. Quantum Inf. Process. 13(6), 1345-1351 (2014)

\bibitem{27} Wang T Y, Cai X Q, Zhang R L.: Security of a sessional blind signature based on quantum cryptograph. Quantum Inf. Process. 13(8), 1677-1685 (2014)

\bibitem{28} Lo, H., Ko, T.: Some attacks on quantum-based cryptographic protocols. Quantum Inf. Comput. 5(1), 41-48 (2005)

\bibitem{29} Gao, F., Guo, F.,Wen, Q., Zhu, F.: Comment on ¡°Experimental Demonstration of a Quantum Protocol for Byzantine Agreement and Liar Detection¡±. Phys. Rev. Lett. 101, 208901 (2008)

\bibitem{30} Zhang, Y., Li, C., Guo, G.: Comment on ¡°Quantum key distribution without alternativemeasurements¡±[Phys. Rev. A 61, 052312 (2000)]. Phys. Rev. A. 63, 036301 (2001)

\bibitem{31} Gao, F., Qin, S.,Wen, Q., Zhu, F.: A simple participant attack on the Bradler-Dusek protocol. Quantum Inf. Comput. 7(4), 329-334 (2007)

\bibitem{32} Wang, T.,Wen, Q., Chen, X.: Cryptanalysis and improvement of a multi-user quantum key distribution protocol. Opt. Commun. 283(24), 5261-5263 (2010)

\bibitem{33} Gao, F., Wen, Q., Zhu, F.: Teleportation attack on the QSDC protocol with a random basis and order. Chin. Phys. B. 17(9), 3189 (2008)

\bibitem{34} Wang, T.,Wen,Q.,Gao, F., et al.: Cryptanalysis and improvement ofmultiparty quantumsecret sharing schemes. Phys. Lett. A. 373(1), 65-68 (2008)

\bibitem{35}  Gao, F., Qin, S., Guo, F., Wen, Q.: Dense-coding attack on three-party quantum key distribution protocols. IEEE J. Quant. Electron. 47(5), 630-635 (2011)

\bibitem{36} Hao, L., Li, J., Long,G.: Eavesdropping in a quantum secret sharing protocol based on Grover algorithm and its solution. Sci. Chin. Phys. Mech. Astron. 53(3), 491-495 (2010)

\bibitem{37} Qin, S., Gao, F.,Wen, Q., Zhu, F.: Improving the security of multiparty quantum secret sharing against an attack with a fake signal. Phys. Lett. A.  357(2), 101-103 (2006)

\bibitem{38}  W¨®jcik, A.: Eavesdropping on the ¡°Ping-Pong¡± quantum communication protocol. Phys. Rev. Lett. 90, 157901 (2003)

\bibitem{39}  W¨®jcik, A.: Comment on ¡°Quantum dense key distribution¡±. Phys. Rev. A. 71, 016301 (2005)

\bibitem{40} Cai, Q.: The Ping-Pong protocol can be attacked without eavesdropping. Phys. Rev. Lett. 91, 109801 (2003)

\bibitem{41} Gao, F., Guo, F.,Wen, Q., Hu, F.: Consistency of shared reference frames should be reexamined. Phys. Rev. A. 77, 014302 (2008)

\bibitem{42} Gao, F.,Wen, Q., Zhu, F.: Comment on: ¡°Quantum exam¡± [Phys. Lett. A 350 (2006) 174]. Phys. Lett. A. 360(6), 748-750 (2007)

\bibitem{43}Gao, F., Lin, S., Wen, Q., Zhu, F.: A special eavesdropping on one-sender versus N-receiver QSDC protocol. Chin. Phys. Lett. 25(5), 1561 (2008)

\bibitem{44} Gao, F., Qin, S., Wen, Q., Zhu, F.: Cryptanalysis of multiparty controlled quantum secure direct communication using Greenberger-Horne-Zeilinger state. Opt. Commun. 283(1), 192-195 (2010)

\bibitem{45} Gisin, N., Fasel, S., Kraus, B., Zbinden, H., Ribordy, G.: Trojan-horse attacks on quantum-key-distribution systems. Phys. Rev. A. 73, 022320 (2006)

\bibitem{46} Deng, F., Li, X., Zhou, H., Zhang, Z.: Improving the security of multiparty quantum secret sharing against Trojan horse attack. Phys. Rev. A. 72, 044302 (2005)

\bibitem{47} Jain N, Anisimova E, Khan I, et al.: Trojan-horse attacks threaten the security of practical quantum cryptography. New J. Phys. 16(12), 123030 (2014)

\bibitem{48} Yang Y G, Sun S J, Zhao Q Q.: Trojan-horse attacks on quantum key distribution with classical Bob. Quantum Inf. Process. 14(2), 681-686 (2014)

\bibitem{49}  Wang, T., Wen, Q.: Security of a kind of quantum secret sharing with single photons. Quantum Inf. Comput. 11(5), 434-443 (2011)

\bibitem{50}  Wang, T., Wen, Q., Zhu, F.: Cryptanalysis of multiparty quantum secret sharing with Bell states and Bell measurements. Opt. Commun. 284(6), 1711-1713 (2011)

\bibitem{51} Tian Y, Chen H, Ji S, et al.: A broadcasting multiple blind signature scheme based on quantum teleportation. Opt. Quan. Elec. 46(6), 769-777 (2014)

\bibitem{52} Kim T, Choi J W, Jho N S, et al.: Quantum messages with signatures forgeable in arbitrated quantum signature schemes. Phys. Scr. 90(2), 025101 (2015)








\end{thebibliography}


\end{document}